\documentclass[twocolumn]{aastex62}

\received{20 July 2020}
\revised{22 October 2020}
\accepted{12 November 2020}
\published{\ldots}
\usepackage{soul}
\reportnum{Accepted to ApJ}

\shorttitle{X-ray and radio bursts from \src}
\shortauthors{Israel et al.}

\begin{document}

\title{X-RAY AND RADIO BURSTS FROM THE MAGNETAR \src}
\correspondingauthor{G. L. Israel}

\author[0000-0001-5480-6438]{G.\,L. Israel}
\affil{INAF--Osservatorio Astronomico di Roma, via Frascati 33, I-00078 Monteporzio Catone, Italy}
\email{gianluca.israel@inaf.it}
\author[0000-0002-8265-4344]{M. Burgay}
\affil{INAF--Osservatorio Astronomico di Cagliari, Via della Scienza 5, I-09047 Selargius, Italy}
\author[0000-0003-2177-6388]{N. Rea}
\affil{Institute of Space Sciences (ICE, CSIC), Campus UAB, Carrer de Can Magrans s/n, 08193, Barcelona, Spain}
\affil{Institut d'Estudis Espacials de Catalunya (IEEC), Carrer Gran Capit\`a 2--4, 08034 Barcelona, Spain}
\author[0000-0003-4849-5092]{P. Esposito}
\affil{Scuola Universitaria Superiore IUSS Pavia, Palazzo del Broletto, piazza della Vittoria 15, I-27100 Pavia, Italy}
\affil{INAF--Istituto di Astrofisica Spaziale e Fisica Cosmica di Milano, via A.\,Corti 12, I-20133 Milano, Italy}
\author[0000-0001-5902-3731]{A. Possenti}
\affil{INAF--Osservatorio Astronomico di Cagliari, Via della Scienza 5, I-09047 Selargius, Italy}
\author{S. Dall'Osso}
\affil{Gran Sasso Science Institute, viale F. Crispi 7, I-67100, L'Aquila, Italy}
\affil{INFN - Laboratori Nazionali del Gran Sasso, I-67100, L’Aquila, Italy}
\author[0000-0002-0018-1687]{L. Stella}
\affil{INAF--Osservatorio Astronomico di Roma, via Frascati 33, I-00078 Monteporzio Catone, Italy}
\author[0000-0001-7397-8091]{M. Pilia}
\affil{INAF--Osservatorio Astronomico di Cagliari, Via della Scienza 5, I-09047 Selargius, Italy}
\author[0000-0002-6038-1090]{A. Tiengo}
\affil{Scuola Universitaria Superiore IUSS Pavia, Palazzo del Broletto, piazza della Vittoria 15, I-27100 Pavia, Italy}
\affil{INAF--Istituto di Astrofisica Spaziale e Fisica Cosmica di Milano, via A.\,Corti 12, 20133 Milano, Italy}
\affil{Istituto Nazionale di Fisica Nucleare (INFN), Sezione di Pavia, via A.\,Bassi 6, I-27100 Pavia, Italy}
\author[0000-0001-9477-5437]{A. Ridnaia}
\affil{Ioffe Institute, 26 Politekhnicheskaya, St Petersburg, 194021, Russia}
\author[0000-0002-7851-9756]{A. Y. Lien}
\affil{Center for Research and Exploration in Space Science and Technology,
NASA Goddard Space Flight Center, Greenbelt, MD 20771, USA}
\affil{Department of Physics, University of Maryland, Baltimore County, 1000 Hilltop Circle,
Baltimore, MD 21250, USA}
\author[0000-0002-1153-6340]{D. D. Frederiks}
\affil{Ioffe Institute, 26 Politekhnicheskaya, St Petersburg, 194021, Russia}
\author[0000-0001-5326-2010]{F. Bernardini}
\affil{INAF--Osservatorio Astronomico di Roma, via Frascati 33, I-00078 Monteporzio Catone, Italy}
\affil{INAF $-$ Osservatorio Astronomico di Capodimonte, Salita Moiariello 16, I-80131 Napoli, Italy}
\affil{New York University Abu Dhabi, Saadiyat Island, Abu Dhabi, 129188, United Arab Emirates}

\def\src{1E\,1547.0--5408}
\def\xte{XTE\,J1810--197}
\def\xmm {\emph{XMM-Newton}}
\def\cxo {\emph{Chandra}}
\def\swift {\emph{Swift}}
\def\sax {\emph{BeppoSAX}}
\def\rxte {\emph{RXTE}}
\def\rst {\emph{ROSAT}}
\def\asca {\emph{ASCA}}
\def\flux {\mbox{erg cm$^{-2}$ s$^{-1}$}}
\def\lum {\mbox{erg s$^{-1}$}}
\def\nh {$N_{\rm H}$}

\newcommand{\comm}[1]{\textcolor{red}{#1}}
\renewcommand{\bf}{\rm}

\begin{abstract}
We report on simultaneous radio and X-ray observations of the radio-emitting magnetar \src\ on 2009 January 25 and February 3, with the 64-m Parkes radio telescope and the \cxo\ and \xmm\ X-ray observatories. The magnetar was observed in a period of intense X-ray bursting activity and enhanced X-ray emission. We report here on the detection of two radio bursts from \src\, reminiscent of Fast Radio Bursts (FRBs). One of the radio bursts was anticipated by $\sim$1\,s (about half a rotation period of the pulsar) by a bright SGR-like X-ray burst, resulting in a $F_{\rm radio}/F_{\rm X}\sim10^{-9}$. Radio pulsations were not detected during the observation showing the FRB-like radio bursts, while they were detected in the previous radio observation. We also found that the two radio bursts are neither aligned with the latter radio pulsations nor with the peak of the X-ray pulse profile (phase shift of $\sim$0.2).
Comparing the luminosity of these FRB-like bursts and those reported from SGR\,1935+2154, we find that the wide range in radio efficiency and/or luminosity of magnetar bursts in the Galaxy may bridge the gap between ``ordinary" pulsar radio bursts and the extragalactic FRB phenomenon.

\end{abstract}

\section{Introduction}
\label{sec:intro}

Among the isolated neutron stars, there is a sub-class of bright X-ray pulsars believed to be powered by their large magnetic fields, of the order of $\sim10^{13-15}$\,G (see \citealt{kaspi17,esposito18} for recent reviews). Among their hallmarks, there are: i) slow rotation periods in the 0.3--12\,s range, ii) X-ray persistent emission modelled by the thermal (0.3--1\,keV) emission of a surface hot spot plus a non-thermal magnetospheric component (power law with $\Gamma\sim$2--4), iii) the emission of X-ray bursts on a wide range of luminosities and timescales ($\approx10^{38-45}$\lum ; ms to minutes), and iv) large X-ray outbursts lasting years.

The discovery of transient pulsed radio emission following intense X-ray outbursts 
of five members of the class \citep{hgb05,camilo06,esposito20} was a bolt in the field. The pulsed radio emission from magnetars is characterised by a flat radio spectral index ($S_{\nu}\propto\nu^{0.5}$) and large variabilities both in flux density and pulse profile \citep[e.g.]{camilo06,kramer07}. 
To further blur the line between magnetars and `ordinary' pulsars, magnetar-like X-ray activity was found in objects with dipolar B fields as low as $6\times10^{12}$\, G \citep{rea10,rea13}, and in pulsars with powerful rotational energy loss rate, such as the radio-quiet PSR\,J1846$-$0258 \citep{gavriil08} and the radio-loud PSR\,J1119$-$6127 \citep{archibald16}.

\begin{table*}[t]
\begin{minipage}{15.5cm}
\caption{Main parameters of the X-ray and radio observations of \src. }
\label{simultab}
\begin{tabular}{@{}ccccccccc}
\hline
Date   & \multicolumn{2}{c}{Start Time (UT)}  & Overlap & $T_{\rm X}/T_{\rm Radio}$ & Flux$_{\rm X}^a$ & \multicolumn{2}{c}{Radio band}  & X-ray $P$\\
        (mm/dd/yy)    & Radio& X-ray &    (hr)  & (hr)    & (erg/cm$^{2}$/s)  & \multicolumn{1}{c}{$\nu$ (MHz)}  & \multicolumn{1}{c}{$\Delta\nu$ (MHz)}         & (s)\\
\hline
\hline
 {\bf 01/23/2009}   &  23:10:11  & --  & --  &  --/1.17     &      --          & 3094.0$^b$ & 1024.0$^b$ &  --  \\
        {\bf 01/25/2009}    & 16:48:22 & {\bf 15:44:01}$^c$ & 1.17  & 3.37/1.17     &    5.74$\pm^{0.02}_{0.41}$          & 2935.5/3094.0$^b$ & 576.0/1024.0$^b$  &   2.07213(2)    \\
        {\bf 02/03/2009}  & 18:28:58 & {\bf 18:23:50}$^d$ &1.51 & 15.68/1.51 & 4.52$\pm^{0.01}_{0.04}$ & 6592.5/6380.5$^b$  & 576.0/256.0$^b$  & 2.072151(2) \\
\hline
\end{tabular}
\medskip\\
$^a$ X-ray flux in units of $10^{-11}$ and in the 0.5--10\,keV energy range. Values are from \citet{bernardini11}.\\
$^b$ DFB observations. 
$^c$ \cxo\, observation. 
$^d$ \xmm\, observation.
\end{minipage}
\end{table*}

A renewed interest for magnetar radio emission comes from its possible connection to Fast Radio Bursts (FRBs). FRBs are bright ($\sim$~Jy) ms-duration transients whose impulsive nature and extreme brightness temperatures imply coherent emission and connect them to compact objects. When the first repeating FRB was observed \citep{spitler14,spitler16} it became clear that at least a selection of FRBs could not be powered by a single explosive event. All these characteristics pointed to a connection with magnetar bursts, possible from very young extra-Galactic magnetars \citep{metzger17,beloborodov17}. Recently, the detection of a double-peaked radio burst simultaneous with a bright SGR-like burst from SGR\,1935+2154 \citep{chime20sgr,bochenek20,mereghetti20,hxmt,ridnaia20} showed for the first time that magnetar bursts can indeed have bright radio counterparts. 

\src\, has a $\sim$\,2.07\,s spin period and a surface dipolar magnetic field of $B_{\rm{p}}\sim 6.4\times10^{19} (P\dot{P})^{1/2} \sim 6.4\times10^{14}$\,G \citep{dib12}. Since its discovery \citep{lamb81,gelfand07}, \src\, has experienced at least three outbursts (in 2007, 2008 and 2009) during which emitted several energetic short bursts \citep{israel10,bernardini11,scholz11,cotizelati20}. Multiple expanding X-ray rings were detected around the source during the 2009 outburst decay, allowing an estimate of the source distance of  $\sim$4.5\,kpc (\citealt{tiengo10}; in the following, we adopt this distance value). 

In this work, we present simultaneous X-ray and radio observations of \src\, performed in 2009, while it was undergoing its brightest outburst. Summary of the results and discussion are presented in \S\ref{sec:results} and \S\ref{sec:discussion}.

\section{Observations and analysis}
\label{sec:obs}

\subsection{X-ray datasets}
\label{sec:X}

On 2009 January 25, \src\ was observed with the S3 CCD of the ACIS camera on board \cxo\ (Table\,\ref{simultab}). The observations were performed in Continuos Clocking mode (time resolution of 2.85\,ms) and standard data processing and reduction were performed with the CIAO software package (v.\,4.10) and the calibration files in CALDB (v.\,4.8). 
We extracted the source events in the 0.3--10 keV energy range from a 3.5\,arcsec region around the source position (RA = 15:50:54.12, Dec = -54:18:24.05, J2000), and the background from a region of the same size.

\begin{figure}
\vspace{2mm}
\hspace{-2mm}\resizebox{.74\textwidth}{!}{\includegraphics[angle=0,width=7.0cm]{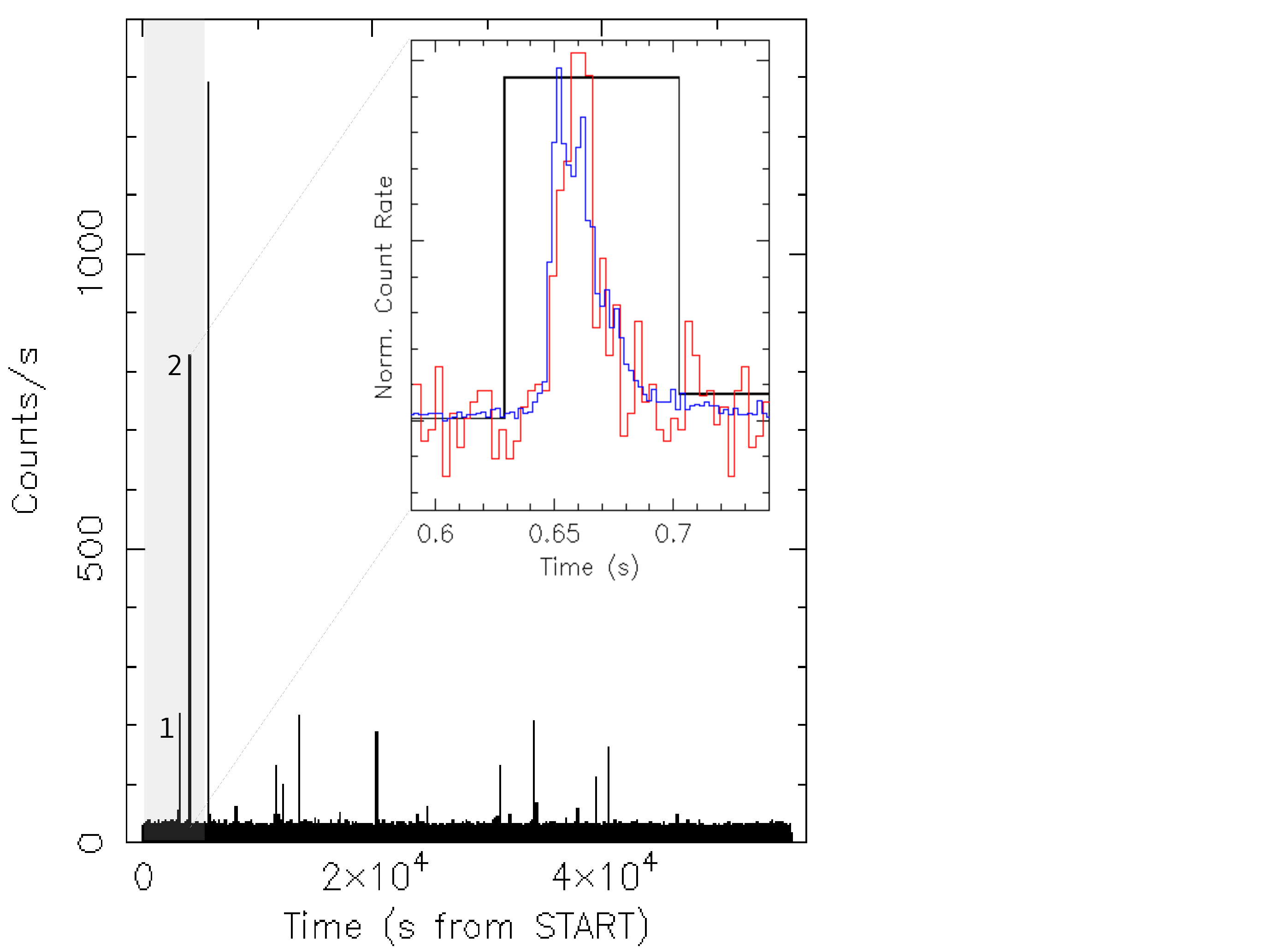}}
\vspace{-7mm}
\caption{\label{fig:timeseries_feb09} 
\xmm\ EPIC-pn 2--10 keV light curve binned at 0.5 s, where the gray area marks the time interval covered by Parkes observations. The detected X-ray bursts within the radio overlap interval are marked as ``1'' and ``2''. The inset shows burst ``2'' as observed by EPIC-pn in the 2--10 keV\,band (black line; time resolution 73.4\,ms), \swift/BAT in the 15--100\,keV band (red line; time resolution of 3\,ms) and KW in the 20--1400\,keV band (blue line; time resolution of 2\,ms). }
\end{figure}

On 2009 February 3--4, \src\ was observed with \xmm\, with all EPIC cameras operating in Full Frame mode (frame time of 73.4\,ms and 2.6\,s, for the pn and MOS, respectively). The data were processed and reduced with standard procedures using the SAS software package. The source photons were accumulated from a circular region with radius of 36$\arcsec$, while the background was estimated from a circle of the same size in the same chip as the source. 
To search for bursts, we extracted the EPIC-pn events from the full detector because
bright magnetar bursts can cause heavy pile-up and other saturation effects that substantially reduce the number of valid events  at the source position.
To study the properties of the persistent emission of the source, the time intervals of the detected bursts (see Figure\,\ref{fig:timeseries_feb09}) were excluded from the analysis by applying an intensity filter (with a negligible reduction of the net exposure time; see Table~\ref{simultab}). 

\src\ also triggered Konus-Wind (KW; \citealt{aptekar95}) during our simultaneous radio/X-ray campaign on 2009 February 3. KW consists of two identical NaI(Tl) scintillation detectors with a 2$\pi$ steradians field of view,
operating from 20\,keV 
to 16\,MeV.
In the triggered mode, activated when the count rate in 
the 80-320 keV band exceeds a $\approx$9$\sigma$ threshold above background, lightcurves with a time resolution up to 2\,ms are recorded in three energy bands,
starting from 0.512 s before the trigger time T0.

{\bf 
At the same epoch also the Burst Alert Telescope (BAT) on board \swift\ \citep{gehrels04}, detected a significant count rate increase coincident in time (and shape) with the KW trigger.
BAT is a coded-mask instrument operating in the 15-150\,keV band with a field of view (FoV) of about 1.4 steradians (half-coded). At the time of the trigger \src\ was outside the BAT FoV, and no source was found in the BAT image. However, high-energy photons can reach the BAT detector from different directions, and even an out-of-FoV source can sometimes cause a count-rate increase in the BAT detectors\footnote{See   \url{https://swift.gsfc.nasa.gov/analysis/bat\_swguide\_v6\_3.pdf.}}.
No spectral information can be obtained in such cases and only timing information are recorded.}


For the timing analysis, the \cxo, \xmm, {\bf KW and \swift} photon arrival times were referred to the barycenter of the Solar System using the JPL-DE405 ephemeris. The epochs of the {\bf bursts} discussed in the following are in barycentric dynamical time (TDB).

\subsection{Radio datasets}
\label{sec:radio}

We observed \src\ with the Parkes 64-m telescope three times during its 2009 outburst: on 2009 January 23 and 25, and on February 3. 
On January 23 data were recorded at a central frequency of 3.1 GHz with the digital filterbank backend DFB3,\footnote{See \url{http://www.srt.inaf.it/media/uploads/astronomers/dfb.pdf}.} over a 1024 MHz bandwidth split into 1-MHz wide channels and were on-line folded with constant period to form 1024 phase bins and 30 s long subintegrations, for a total of 1.2 hr. On January 25, in parallel with the DFB3, we also acquired data with the analogue filterbank AFB in search mode, over a total bandwidth of 576 MHz (centered at 2.9 GHz) split in 3-MHz wide channels. AFB data were 1-bit digitized every 1 ms, while DFB3 data were folded on-line with the same parameters and duration as the previous observation, providing total overlap with X-ray data. For the third observation on February 3, we collected data at a central frequency of 6.6 GHz for 1.5 hr beginning at 18:28:58, i.e. about 5 minutes after the start of the X-ray observation. Data were collected simultaneously with the AFB using 3-MHz-wide channels covering a total bandwidth of 576 MHz and with the DFB3 using 0.5-MHz-wide channels and a total bandwidth of 256 MHz centered at 6.4 GHz. AFB data were 1-bit sampled every 1 ms, while DFB3 data were folded on-line with the same parameters adopted for the previous observations (1024 phase-bins, 30-s subintegrations).

\section{Results}
\label{sec:results}

\subsection{X-ray results}

By means of a phase-coherent timing analysis, we measured a spin period of $P =  2.07213(2)$\,s in the \cxo\ data set, and $P = 2.072151(2)$\,s in the \xmm\ one (1$\sigma$ confidence level).  We checked that these values are in agreement with the long-term phase-coherent timing solution presented in \citet{bernardini11}. 
Both the pulse shape and the pulsed fraction (PF, defined as the semi-amplitude of a sinusoidal function divided by the source average count rate) changed significantly between the two X-ray observations: from an almost sinusoidal to a double-peaked pulse profile, and from $\rm PF = 10\pm1$\,\% to $15\pm1$\% (in the 0.5--10\,keV interval).


The search for bursts was performed on the 2--10\,keV light curve of the full EPIC-pn detector, with bin time of 0.5 s. We label as a burst every bin 
with a probability $<$0.001 of being a Poissonian fluctuation of the average count-rate, considering the number of time bins of the light curve as the number of independent trials. Based on this definition, 
we detected 11 bursts in the \xmm\, X-ray lightcurve and only two in the interval covered also by the Parkes observation. The TDB time of the latter two bursts (``1'' and ``2'' in Fig.\,\ref{fig:timeseries_feb09}) are 19:17:10.5 and 19:31:28.5. 
{\bf The 0.5 s time bin associated to the two bursts contain 90 and 374 pn events, respectively.} 

During the peak of burst ``2'', some of the pn quadrants, consisting of three CCDs sharing the same electronics, were fully saturated  and registered no valid events, making it impossible to evaluate the burst fluence. 
On the other hand, burst ``2'' was detected in a single 2.6 s frame in both the MOS cameras. In this case, the burst is heavily piled-up, with no events detected in the PSF core,
but a reliable spectrum could be extracted from a 2--5 arcmin annular region{\bf, containing 120 and 122 counts in MOS1 and MOS2, respectively, in the 2--10 keV energy range}.

Burst ``2'' was also seen by \swift\ BAT (TDRSS ObsID 00341964000) even though the source was out of the nominal FoV at the time of the event. No spectral information are available but it provided us with an accurate peak barycentred time of 19:31:28.66~TDB. 

Furthermore, 
burst ``2'' 
triggered KW at $ T_{0,\,\mathrm{KW}}$ = {\bf 19:31:28.646~TDB (with an accuracy of few ms).}
The EPIC-pn, BAT and KW light curves are shown in Fig.\,\ref{fig:timeseries_feb09}.
 
Spectral information could be derived from the KW data, from 16 to 80\,ms after the rise of the burst. {\bf This spectrum contains 400 background-subtracted counts in the 20--300 keV energy band.} 
The lightcurves in the 20--80 keV and 80--350 keV bands show no evidence of spectral changes along the burst evolution. We therefore performed a joint analysis of the KW and MOS spectra by applying to the model of the KW spectrum a cross-normalization factor {\bf accounting for the instrumental dead-time and}
the fraction of burst background-subtracted counts detected in this 64 ms time interval. We obtained a good fit ($\chi^2_{\nu}=1.6$ for 16 d.o.f.) with a double blackbody model, with photoelectric absorption fixed at \nh = $4.2\times10^{22}$ cm$^{-2}$ \citep{pintore17}, obtaining the following best-fit parameters:\footnote{\bf To evaluate the blackbody radii, a burst duration of 20 ms was assumed.} {\bf $kT_1 = 4.1\pm0.2$\,keV and $R_1=65\pm4$\,km, $kT_2 = 14.3\pm0.8$\,keV and $R_2=5.5\pm0.8$\,km.}  From this model, we derive a bolometric fluence of {\bf 2.6$\times$10$^{-6}$ erg cm$^{-2}$. The observed fluence in the 2--10 keV and 20--100 keV is $3.1\times10^{-7}$\,erg\,cm$^{-2}$ and $1.5\times10^{-6}$\,erg\,cm$^{-2}$,} respectively. A much worse fit {\bf ($\chi^2_{\nu}=2.2$} for 17 d.o.f.) was obtained by adopting an absorbed cutoff power-law model.
 
 Also burst ``1'' was detected by the MOS cameras. Due to its lower fluence, we could safely extract its spectrum from an annulus with a smaller inner radius (20 arcsec), {\bf containing 35 counts in MOS1 and 37 counts in MOS2}. Assuming the same best-fit spectral model as burst ``2'', its fluence in the 2--10 keV energy band is only 
 4.5$\times$10$^{-9}$
 erg cm$^{-2}$.

\begin{figure}
\hspace{-6mm}\vspace{-1mm}\resizebox{.5\textwidth}{!}{\includegraphics[angle=-90]{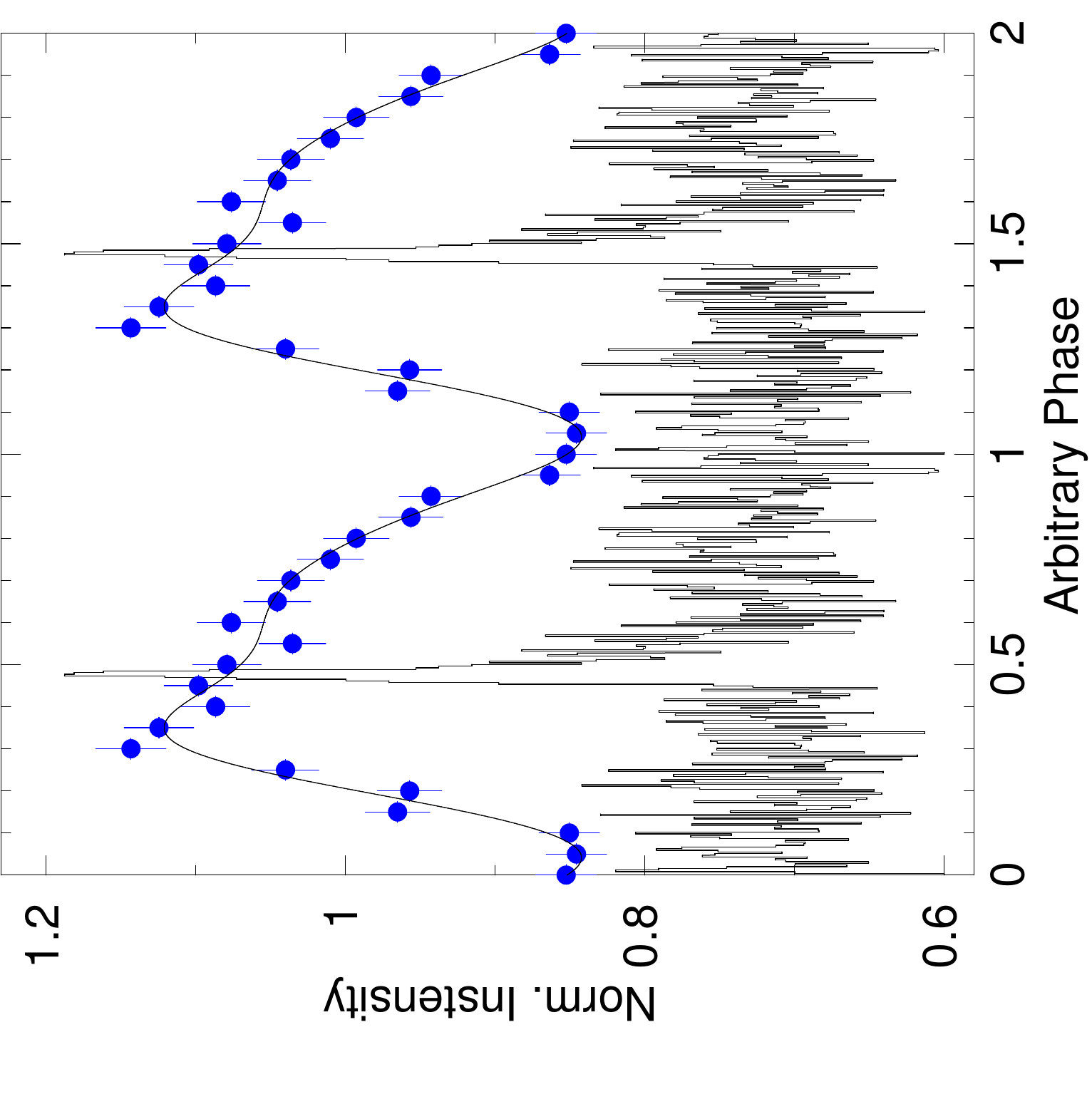}}
\caption{
X-ray and radio pulse profiles of \src\, from the \cxo\ ACIS-S (0.5--10 keV) and 3.1 GHz Parkes light curves obtained on 2009 January 25. Superimposed to the X-ray folded light curve is the best fit obtained with the sum of sinusoidal functions.
} \label{fig:xradio_feb09}
\end{figure}

\begin{figure*}
\resizebox{1.\textwidth}{!}{\includegraphics[angle=0,width=6cm]{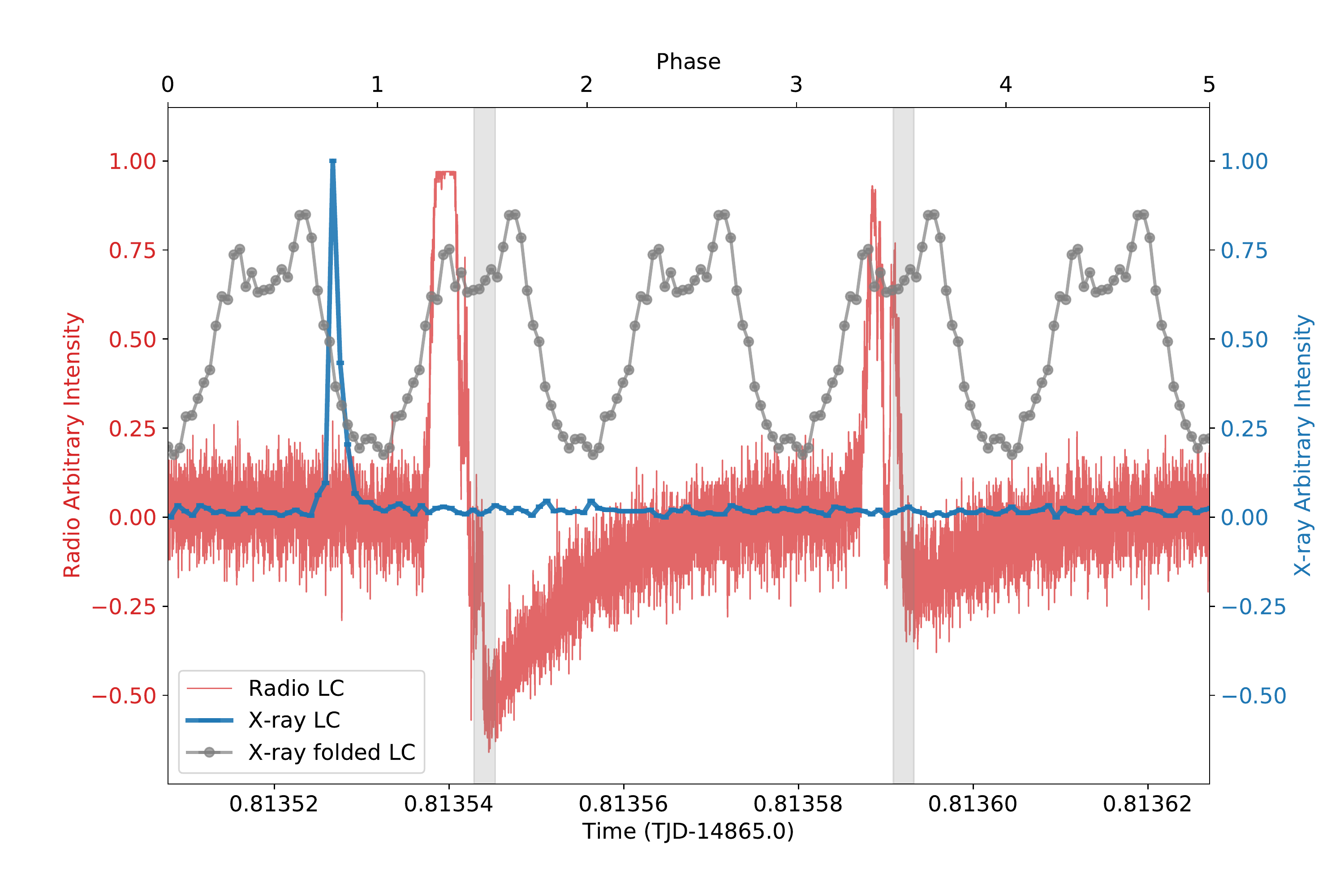}}
\vspace{-10mm}
\caption{\label{fig:allphasestemp} 
X-ray and radio simultaneous observations of \src\, performed on 2009 February 3, around the time of the brightest X-ray burst (burst ``2'' in Fig.\,\ref{fig:timeseries_feb09}) and the two radio pulses (pulse A and B). The blue line is the burst lightcurve, while the grey line is the X-ray folded light curve using the \xmm\, dataset. The Parkes simultaneous radio light curve is shown in red. The grey shaded areas are the phase intervals of the expected peak of the radio pulse profiles extrapolated from the 2009 January 25 Parkes observation (see also  Figure\,\ref{fig:xradio_feb09}). Note that the flat top of the first radio peak (pulse A), and the drop of the intensity of the radio signal below the average noise level following both pulses, are artifacts caused by the saturation of the backend.}
\end{figure*}

 \subsection{Radio results}
 
We first searched for radio pulsations during the three Parkes observations. No radio pulsations were detected from \src\ in our first observation, taken on January 23rd, down to a flux density limit of $\sim 0.06$ mJy. The source was again visible as a radio pulsar on January 25th \citep{burgay09}. As expected, the rotational ephemeris available in the literature did not satisfactorily phase align the radio data. A local timing solution obtained from the X-ray data \citep{bernardini11} was used to correctly fold off-line the radio data. In the third observation on February 3rd, no pulsations were detected, but only two single pulses/bursts were observed, at epochs 19:31:29.82 (pulse A) and 19:31:34.05 (pulse B) {\bf TDB}. The reported times are 
also corrected for the dispersion delay (using the dispersion measure DM = $830 \pm 50\, pc/cm^3$, from \citealt{camilo07a}) at the observing frequency. The error on the DM results in an uncertainty in the bursts arrival times of 4.4 ms. All times are reported at infinite frequency.

 Pulse A was so strong to saturate for about 0.2 seconds the AFB backend. This can be clearly seen from the red curve in Fig \ref{fig:allphasestemp}, where pulse A shows a flat top and the baseline following it shows a significant distortion. Pulse B was also likely very close to the saturation level, as can be inferred from the value of its highest point and from the (smaller than for pulse A) distortion of the baseline. For pulse A the time of arrival reported above refers to the mid-point of the saturated portion, while for pulse B to the highest point in the profile.

Taking into account the 1-bit sampling, the 1-ms sampling time, and the frequency width of the channels in use, we can place a lower limit for the flux density necessary to saturate the backend. Using the receiver's system temperature and gain (55 K and 0.71 K/Jy respectively\footnote{https://www.parkes.atnf.csiro.au/observing/documentation/}), we obtain a peak flux density lower limit for the pulse A of $\gtrsim 1$ Jy. Given the 0.2 s duration of the saturation, this corresponds to a lower limit in the pulse fluence of $\gtrsim 200$ Jy ms. 
 
Thanks to the simultaneous observations obtained with the DFB3, we can also give an approximate estimate of the actual peak flux density of the saturated pulse. Even though the data were folded on-line, in fact, the two pulses fall in two separate, consecutive, 30-s long sub-integrations. The signal-to-noise ratio of pulse A (averaged with $\sim 15$ `silent' rotations) is, in the DFB3 data, three times larger than that of pulse B, also smoothed by the noise resulting from $\sim 15$ `silent' rotations. Since the pulse B is very close to saturate the AFB, we can tentatively conclude that pulse A must be approximately three times stronger than the saturation limit for the AFB. We can hence estimate a fluence of $\sim 0.6$ kJy ms.

To align in phase the X-ray and radio observations of the bursts, we first measured the times of arrival (ToAs) of a fiducial point of the radio profiles (the peak or, in the case of the saturated pulse, the mid point of the saturation plateau).  For the two single pulses the times were obtained directly from the time series, barycentered and corrected for the dispersion delay at the observed radio frequency using \textsc{prepdata}, from the \textsc{presto} package\footnote{\url{https://www.cv.nrao.edu/~sransom/presto/}}. On the other hand, the ToA of the peak of the folded profile of the pulsations observed on January 25, was measured by cross-correlation with a synthetic profile template using \textsc{pat} from the \textsc{psrchive} package\citep{vanstraten+2012}, and then barycentered and corrected for the dispersion delay using the \textsc{general2} plugin in \textsc{tempo2}\footnote{\url{https://bitbucket.org/psrsoft/tempo2/src/master/}} \citep{hobbs06}. A local set of ephemerides obtained from the simultaneous X-ray observations was used in the process. The same ephemerides were used to determine the fractional part of the absolute phase of the radio pulses, from which we obtained the phase offset between the selected bin in the radio profiles and the beginning of the first bin of the X-ray light curve.

The results are shown in Fig.\,\ref{fig:xradio_feb09} and Fig.\,\ref{fig:allphasestemp} where the total X-ray folded light-curve is almost aligned  with the radio pulses. This result is at variance with that obtained by \citet{halpern08}, who found $\sim$0.2 cycles phase shift between the X-ray and radio profiles {\bf(with the X-rays peak preceding the radio one)}. This is not surprising because the source is known to vary its profile shape and hence also the exact position of its main peak {\bf \citep{bernardini11,camilo07a}}. Moreover, the single pulses (even for otherwise ordinary pulsars) are often seen to {\bf wander around} the integrated peak profile {\bf with notable exceptions (as an example see the case of XTE J1810-197; \citealt{camilo07b})}.

 \subsection{Radio/X chance coincidence probability}

We estimated the probability of having two radio pulses occurring by chance within $\sim 1.2$ s (about half a rotation cycle) of the 
two X-ray bursts detected in the window of radio/X-ray simultaneity on February 3. The 0.5\,s bin time of
the X-ray light curve 
corresponds to the zero height width of the saturated radio pulse
and yields to ${\cal N}\approx$ 10\,800 bins in the time interval of interest.
As a conservative approximation, we assume the probability of chance coincidence as the ratio between $N_{\rm c}$,  the total number of configurations in which one random radio pulse may follow one X-ray burst by $\le 3$ bins, and ${\cal N}$, the total number of bins. We obtain $N_{\rm c} = 12$ as the number of trials (2 radio pulses) times the number of  targets (2 X-ray bursts) times the number of valid bins per trial (3), which implies a (maximum) chance coincidence probability of 1.1$\times$10$^{-3}$ (3.3$\sigma$). Similarly,  we obtain a probability of 7.3$\times$10$^{-4}$ (3.5$\sigma$) by considering the two-bin delay ($\lesssim$1\,s) between the X-ray burst peak and the start of the radio pulse.

\section{Discussion}
\label{sec:discussion}

We performed two X-ray and three radio observations of \src\, during its 2009 burst active phase.~In our 1.5\,hr-long 2009 February 3rd observing campaign we detected two bright radio pulses/bursts.~In~particular, the first 
event (pulse A), with fluence $F \sim 0.6$\,kJy\,ms and width $\sim 200$ ms, occurred about 1\,s (half a rotation cycle) after a very bright X-ray burst (burst~``2'') showing a profile with two 
peaks separated by 10 ms. The X-ray burst had a bolometric fluence of 1.3$\times$10$^{-6}$~erg~cm$^{-2}$ and width of $\sim 50$ ms. The radio bursts are neither aligned with the radio pulsations detected six days before in a previous radio observation, nor with the X-ray pulsations, presenting a $\sim$ 0.2 phase shift with respect~to both. These findings, along with the lack~of ``normal"~radio pulsations during {\bf our 1.5 hr-long} observation, 
strongly suggest a close connection of the~X-ray and radio bursts.
This new and peculiar magnetar phenomenology assumes additional relevance in the context of the recently detected FRB-like bursts from SGR~J1935+2154  \citep{chime20sgr,bochenek20,mereghetti20,ridnaia20,tavani20b,hxmt}. 
Two bright radio bursts were detected from SGR~J1935+2154, with a total fluence of $\sim 700$~kJy~ms at 600 MHz. The second of them was also independently observed by STARE2 at 1.5 GHz, with an estimated fluence above 1.5~MJy~ms \citep{bochenek20}.
Furthermore, the reported radio bursts show a steep spectral index, as do FRBs. In our case, the spectral properties of the radio bursts cannot be constrained as we only observed at 6.6 GHz.

In the case of SGR~J1935+2154, {\bf despite a slight misalignment of the X-ray peaks as observed by different instruments \citep{mereghetti20,ridnaia20,hxmt}, the radio burst {\bf appears} 
to lead the SGR-like X-ray burst by no more than $\sim$ 8\,ms}, while in \src\ it is the X-ray burst that leads the first radio burst by $\sim 1$~s. 

Comparing the SGR-like X-ray {\bf bursts}  
of SGR~J1935+ 2154 and \src, as detected by KW data, both were in line in {\bf terms of} energy and duration with what usually observed from their respective sources. The spectrum{\bf, in the case of  SGR~J1935+2154,} was harder than {\bf in typical bursts from that source and was indeed among the top 2\% hardest magnetar bursts ever detected by KW \citep{ridnaia20}.} On the other hand, {\bf the softer  spectrum of} \src's burst 
{\bf was very typical within the magnetar population.}

No further simultaneous X-ray and radio bursts were detected despite the continued X-ray activity of SGR~J1935+2154, {\bf while} several fainter radio bursts were reported \citep{fast20atel,kirsten20}. {\bf The source entered a new radio active phase in early October 2020, characterized by one bright burst of fluence $\sim 900$ Jy~ms observed by CHIME, followed by many fainter bursts below 50~Jy~ms observed by both CHIME and FAST \citep{good20,pleunis20,zhu20}. FAST pulses were aligned in phase with the detected radio pulsation of the source.}

The energy released in the brightest radio bursts of SGR~J1935+2154 and \src\ is $3\times 10^{34}$~erg and $8.4\times 10^{30}$~erg, with a radio-to-X-ray ratio $E_r/E_X \sim 10^{-5}$ and 
$\sim 10^{-9}$, respectively. In both cases these were brighter than standard magnetar radio single peaks {\bf (below a few Jy; e.g. \citealt{camilo06})}, yet ``mild'' compared to FRBs, {\bf although the latter are known to span a wide range of energies \citep{chime20sgr,bochenek20}.}  
For the giant flare of SGR~1806--20 \citep{palmer05} the most stringent limits to the ratio of radio versus X-ray emitted energy were set down to $\lesssim 10^{-8}$ \citep{tendulkar16}. On the other hand, in the case of the better constrained FRB repeaters with simultaneous X/radio observations \citep{scholz17,pilia20} these limits are of the order of $\geq 10^{-9} - 10^{-8}$. According to {\bf \cite{crl20}, who derive constraints on $E_r/E_X$ in FRBs from past surveys, $E_r$ should be of order $\sim 10^{-5} E_X$, larger than our detection for \src.}

The wide range in radio efficiency and/or luminosity of magnetar bursts suggests that these sources may bridge the gap between ``ordinary" pulsar radio bursts and the extragalactic FRB phenomenon (as also suggested in \citealt{chime20sgr,bochenek20,kirsten20}). 
The detection of energetic radio bursts simultaneous with regular X-ray bursts {\bf is indicative of what already pointed by previous studies \citep{cunningham19, Casentini2020, guidorzi20, scholz20}, i.e.}
that non-detections of high-energy emission from FRBs can be attributed to limiting telescopes sensitivities rather than intrinsic inhibition of the phenomenon . 

The bursts reported in this work appear to be magnetospheric in origin, given their relatively good phase alignment with the rotation of the magnetar. 
\cite{lyutikov20} propose that magnetospheric reconnection events can explain the energetics and the apparent simultaneity of the X-ray and radio bursts of SGR~J1935+2154. 
In the case of \src, however, the total energy in the radio burst may still be consistent with rotation-powered emission, given its $\sim 0.5$\,s duration and the estimated spin-down luminosity of the source, $\sim 10^{35}$ erg s$^{-1}$. \citep{rea12, lyutikov16, esposito20}.

On the other hand, according to \cite{lkz20}, an explosion at the surface of the neutron star would produce both the X-ray burst and the radio burst, by propagating through the crust towards the polar cap in the form of Alfven waves. In this scenario the emission is magnetospheric and it arises naturally that not all X-ray bursts generate an FRB-like radio burst, as it is observed for both SGR~J1935+2154 \citep{borghese20} and \src. However, the radio emission is expected to lag the X-ray burst, albeit slightly, contrary to the observations in \src.

Moreover, both this and the external models \citep{margalit20} require {\bf a ratio} $L_r/L_X >  10^{-5}- 10^{-4}$, which is in tension with our findings and with observations of the SGR 1806-20 giant flare.

{\bf The picture emerging from our detection of radio and X-ray bursts from \src, as well as from the most recently observed bright and faint radio and X-ray bursts from SGR~J1935+2154 \citep{chime20sgr,bochenek20,fast20atel,kirsten20}, is that there exists a continuum of magnetar radio burst energies, which might at times look like FRBs and at others be much closer to typical radio pulsar single pulse phenomenology (such as Rotating Radio Transients or Giant Pulses, \citealt{mclaughlin04, lyutikov20}). On the other hand, the X-ray counterparts of such radio bursts are not yet easily predictable, and may depend on specific parameters of the burst, most of which are still not fully understood.}

\acknowledgments
GLI and LS acknowledge funding from ASI-INAF agreements I/037/12/0 and 2017-14-H.O. GLI and AT acknowledge financial support from the Italian MIUR PRIN grant 2017LJ39LM. MB, AP, and LS acknowledge funding from the grant ``iPeska'' (INAF PRIN-SKA/CTA; PI Possenti). NR is supported by the ERC Consolidator Grant ``MAGNESIA" (nr.817661), and by grants SGR2017-1383 and PGC2018-095512-BI00. We acknowledge the support of the PHAROS COST Action (CA16214).

\facilities{\xmm, Parkes, Chandra, \swift, Konus-Wind}
\software{FTOOLS \citep{blackburn95}, XSPEC \citep{arnaud96}, XMM-SAS \citep{gabriel04}}


\end{document}